
\magnification=1200
\def\tr#1{{\rm tr #1}}
\def\d{\partial}
\def\f#1#2{{\textstyle{#1\over #2}}}

\def\next{\hfil\break\noindent}
\def\R{{\bf R}}
\def\C{{\bf C}}
\def\ref#1{[#1]}
\font\title=cmbx12

{\title
\centerline{The initial singularity in solutions of the Einstein-Vlasov
system}
\centerline{of Bianchi type I}}

\vskip .5cm\noindent
Alan D. Rendall
\next
Institut des Hautes Etudes Scientifiques
\next
35 Route de Chartres
\next
91440 Bures-sur-Yvette
\next
France

\vskip 2cm\noindent
{\bf Abstract}

The dynamics of solutions of the Einstein-Vlasov system with
Bianchi I symmetry is discussed in the case of massive or massless
particles. It is shown that in the case of massive particles the
solutions are asymptotic to isotropic dust solutions at late
times. The initial singularity is more difficult to analyse.
It is shown that the asymptotic behaviour there must be one of
a small set of possibilities but it is not clear whether all
of these possibilities are realized. One solution is exhibited in
the case of massless particles which behaves quite differently near
the singularity from any Bianchi I solution with perfect fluid as
matter model. In particular the matter is not dynamically negligeable
near the singularity for this solution.

\vskip 2cm\noindent
{\bf 1. Introduction}

The simplest of all anisotropic cosmological models are those of
Bianchi type I. They are the spacetimes which admit a
three-dimensional abelian symmetry group whose orbits are spacelike.
(For general information on Bianchi models see \ref1.) Just how simple
their dynamics is depends significantly on the nature of the matter
content of the spacetime.  For a perfect fluid with a linear equation
of state it has been known for a long time how to analyse the dynamics
\ref{2,3}. For a non-interacting mixture of two fluids with linear
equations of state the time evolution is also well understood and is
asymptotic near the singularity and at large times to that of a single
fluid \ref4. The case of a fluid with nonlinear equation of state is
discussed in an appendix to the present paper. The dynamics does not
differ much from the picture in the linear case.  When a magnetic
field is added to the fluid things are already more complicated. In
fact, as was shown by Collins \ref5, a Bianchi type I model with fluid
and magnetic field resembles a model of the more complicated Bianchi
type II with fluid alone. It is also interesting to note that models
of type VI${}_0$ with perfect fluid and a magnetic field have a
dynamical behaviour resembling the notoriously complicated
\lq Mixmaster\rq\ behaviour of Bianchi type IX models \ref6. Thus changing
the matter model can have effects on the complexity of the dynamics
comparable with those encountered when passing to more general
symmetry types.

A matter model for which the details of the global dynamics of Bianchi type
I spacetimes has not previously been studied mathematically is the
collisionless gas, described by the Vlasov equation. The only general facts
which are known are that, with an appropriate choice of time orientation,

\noindent
(i) the spacetime is future geodesically complete (when maximally extended
towards the future)
\next
(ii) there is a crushing singularity in the past where, except in the
vacuum case, the curvature invariant $R_{\alpha\beta}R^{\alpha\beta}$
tends to infinity

\noindent
These fundamental facts were proved in \ref7, where it was shown
that they hold for any Bianchi type other than IX and for a general
class of matter models. The aim of this paper is to refine (i) and
(ii) in the case of Bianchi type I symmetry and matter described by the
Vlasov equation so as to get more detailed information about the
asymptotics of the expanding phase and the nature of the initial
singularity. An aspect of the situation which makes this more
difficult than in the case of many other matter models is that
for general initial data it is not possible to derive an
explicit closed system of ordinary differential equations which
describes the dynamics. This is because certain integrals which
occur can not be evaluated. In one special case, where massless
particles are considered and the initial phase space density has
the form of the characteristic function of a ball, these integrals
have been computed by Lukash and Starobinski\ref8. However, the
explicit expressions they obtain are sufficiently complicated that
they do not seem to make a rigorous analysis of the global
dynamics any easier. On the other hand they would probably be
useful for numerical calculations since they would allow
costly numerical evaluation of integrals to be avoided.

The dynamics at late times of the models with massive particles can be
described precisely.  All solutions become isotropic and can be
approximated by dust solutions in this limit (Theorem 5.4).  On the
other hand, the results of this paper do not give a complete picture
of the dynamics near the initial singularity of the spacetimes being
studied. They merely reduce the possible types of asymptotic behaviour
to a small number of alternatives. Improving on this is likely to
require new techniques.  These results leave open the possibility that
Bianchi I spacetimes with a matter content described by kinetic theory
may show complicated oscillatory behaviour and thus be very different
from those with other types of matter content studied up to now.  The
mechanism which allows for this complexity is simply the presence of
anisotropy in the pressure which may respond to changes in the
geometry. It may be that the only reason that the dynamics is so
simple in the case of a perfect fluid is that this mechanism is
excluded by a special symmetry assumption (the isotropy of the
pressure). The one conclusion which emerges and which applies to all
solutions considered here is that the ratio of the mean pressure to
the energy density tends to one third as the singularity is
approached. This means that in a certain weak sense the dynamics for
particles of unit mass is approximated near the singularity by that
for massless particles. For this reason both cases are often
considered together in the following, although the main emphasis is on
the case $m=1$.

The results will now be summarized. There are broadly speaking two
possible types of asymptotic behaviour of solutions of the
Einstein-Vlasov system with Bianchi I symmetry near the singularity.
They will be referred to as convergent and oscillatory. Let
$\lambda_i$ denote the eigenvalues of the second fundamental form of
the homogeneous hypersurfaces. Then the mean curvature of the
homogeneous hypersurfaces is given by
$H=\lambda_1+\lambda_2+\lambda_3$. Let $p_i=\lambda_i/H$.  In the
convergent type the $p_i$ tend to limits as the singularity is
approached. There are three different cases, depending on these
limiting values. The first case is that where the limiting values are
$(\f13,\f13,\f13)$.  The well known homogeneous and isotropic solutions
of the Einstein-Vlasov system\ref9 are of this type. The second is
that the limiting values are $(0,\f12,\f12)$ or some permutation
thereof. The existence of solutions of this kind in the case of
massless particles is shown in Section 6.  These limiting values
of the Kasner exponents are not realized by any Bianchi type I
spacetime when the matter model is a perfect fluid (see appendix).
The third is that the
limiting values satisfy the Kasner relation $p_1^2+p_2^2+p_3^2=1$.
Any solution for which one of the eigenvalues becomes negative at some
time has this asymptotic behaviour and so there are plenty of
examples.  This is proved in Theorem 5.1. Note that the special case
of this result when two of the $p_i$ are equal is closely related to
the homogeneous special case of a result of Rein\ref{10} for plane
symmetric spacetimes.  In the oscillatory type the $p_i$ undergo
infinitely many oscillations, in a sense which will now be specified.
There are two cases to be considered, according to whether two of the
eigenvalues are always equal or not. Consider first the case where two
eigenvalues are equal and suppose without loss of generality that
$\lambda_2=\lambda_3$. Associate to any solution a string of symbols
(which may be finite or infinite, depending on the solution) as
follows. Moving backwards from some fixed time add an $x$ to the
string each time that $\lambda_1-\lambda_2$ changes from being $\le 0$
to being $>0$ and add a $y$ each time it changes from being $\ge 0$ to
being $<0$. That this makes sense follows from the fact, proved in
Section 2, that the set of times where $\lambda_1=\lambda_2$ can have
no limit point unless this equality holds at all times.  Thus a finite
time interval can only contribute a finite number of symbols. The
solution is said to undergo infinitely many oscillations if the
resulting string of symbols is infinite. Similarly, if there is some
time where all eigenvalues are distinct then a string of symbols is
associated to the solution by adding $x$, $y$ or $z$ each time
$\lambda_1$, $\lambda_2$ or $\lambda_3$ respectively becomes strictly
larger than the other two eigenvalues.

Unfortunately it could not be shown whether any oscillatory
solutions exist. If they did then the behaviour of Bianchi I models
with kinetic theory as matter model would be much more complicated
than in the case of a perfect fluid. If it could be shown that
they existed the question would remain whether the sequences of
symbols they produce have some regularity or whether they are
chaotic.

To each type of solution discussed above corresponds a characteristic
behaviour of the pressures. The solutions considered in the following
all have diagonal energy-momentum tensors and so three pressures $P_i$
are defined by three diagonal components. The quantities
$R_i=P_i/\rho$ must have a sum which converges to unity at the
singularity. When the limiting values of the $p_i$ are
$(\f13,\f13,\f13)$ or $(0,\f12,\f12)$ then the limiting values of the
$R_i$ are $(\f13,\f13,\f13)$ or $(0,\f12,\f12)$ respectively. When the
sum of the squares of the limiting values of the $p_i$ is equal to
unity then the $R_i$ tend to $(0,\f12,\f12)$ or a permutation thereof,
unless one $p_i$ has limiting value zero. In the latter case the $R_i$
tend to $(1,0,0)$ or a permutation thereof.

The paper is organized as follows. In Section 2 some basic facts
about the solutions are collected. Section 3 is concerned with
a simplified system which in some cases models the asymptotic
behaviour of the solutions of the original system. Some estimates
for the pressures are derived in Section 4. Section 5 contains the
main results. Section 6 contains proofs of the existence or
non-existence of solutions with certain kinds of asymptotic behaviour.

\vskip .5cm
\noindent
{\bf 2. Basic facts}

The Einstein-Vlasov system for particles of mass $m\ge 0$ can be
written in the following form in the case of Bianchi type I symmetry:
$$\eqalignno{ &-k_{ij}k^{ij}+H^2=16\pi\rho&(2.1) \cr
&T_{0i}=0&(2.2)                                  \cr
&\d_t g_{ij}=-2k_{ij}&(2.3)                      \cr
&\d_t k_{ij}=Hk_{ij}-2k_{il}k^l_j-8\pi T_{ij}-4\pi\rho g_{ij}
+4\pi\tr Tg_{ij}&(2.4)                           \cr
&\d f/\d t+2k^i_jv^j\d f/\d v^i=0&(2.5)          \cr
&\rho=\int f(t,v^k)(m^2+g_{rs}v^rv^s)^{1/2}(\det g)^{1/2} dv^1 dv^2 dv^3
&(2.6)                                           \cr
&T_{ij}=\int f(t,v^k)v_iv_j(m^2+g_{rs}v^rv^s)^{-1/2}(\det g)^{1/2} dv^1
dv^2 dv^3 &(2.7)}$$
Here $g_{ij}$ is the induced metric of the homogeneous hypersurfaces,
$k_{ij}$ is the second fundamental form, $f$ is the phase space
density of particles, $\rho$ is the energy density, $T_{0i}$ and
$T_{ij}$ are components of the energy-momentum tensor and $H$ is the
mean curvature $g^{ij}k_{ij}$.  With the exception of $f$ all these
quantities depend only on the time coordinate $t$. This time
coordinate is Gaussian i.e. it is constant on each homogeneous
hypersurface and defines a parametrization by proper time when restricted
to any geodesic normal to these hypersurfaces. The spacetime metric is of
the form
$$ds^2=-dt^2+g_{ij}(t)dx^idx^j\eqno(2.8)$$
In the following it is always assumed, when talking about solutions
of (2.1)-(2.7), that the function $f(t,v)$ is non-negative and
has compact support for
each fixed $t$. It is assumed that $f$ is $C^1$ except in Section
3, where $f$ may be a distribution. The case of primary interest here
is the case $m=1$. Since, however, solutions of (2.1)-(2.7) with
$m=1$ resemble solutions with $m=0$ close to the singularity, it
is useful to also allow the case $m=0$ from the beginning.

For a given Bianchi I geometry the Vlasov equation can be solved
explicitly. (This is not possible for the other Bianchi types.
The reason is explained in \ref{11}.) The result is that if $f$
is expressed in terms of the covariant components $v_i$ then it is
independent of time. This means that if $t_0$ is some fixed time and
$f_0(v_i)=f(t_0,v_i)$ then (2.6) can be rewritten as
$$\rho=\int f_0(v_i)(m^2+g^{rs}v_rv_s)^{1/2}(\det g)^{-1/2}dv_1dv_2dv_3
\eqno(2.9)$$
and (2.7) can be rewritten in a similar way. The explicit solution
allows certain special subclasses of solutions of (2.1)-(2.7)
to be identified. The first of these will be referred to as
reflection-symmetric and is defined by the conditions that
$$f_0(v_1,v_2,v_3)=f_0(-v_1,-v_2,v_3)=f_0(v_1,-v_2,-v_3).\eqno(2.10)$$
and that the initial values of $g_{ij}$ and $k_{ij}$ are diagonal.
Equation (2.10) implies that $T_{ij}$ is diagonal. It then
follows from (2.3) and (2.4) that $g_{ij}$ and $k_{ij}$ are
always diagonal. The second case, which will be referred to
as LRS (locally rotationally symmetric) is obtained by
supplementing (2.10) by the conditions that
$$f_0(v_1,v_2,v_3)=F((v_1)^2+(v_2)^2,v_3)\eqno(2.11)$$
for some function $F$
and that $g_{11}=g_{22}$, $k_{11}=k_{22}$ initially. Equation
(2.11) implies that $T_{11}=T_{22}$ and it follows from
(2.3) and (2.4) that $g_{11}=g_{22}$ and $k_{11}=k_{22}$
everywhere. A solution will also be called LRS if it
satisfies the definition obtained from that just given
by a permutation of the indices $1,2,3$. A solution will
be called isotropic if
$$f_0(v_1,v_2,v_3)=F((v_1)^2+(v_2)^2+(v_3)^2)\eqno(2.12)$$
and if $g_{ij}$ and $k_{ij}$ are proportional to $\delta_{ij}$
on the initial hypersurface and hence everywhere. If a solution
satisfies the conditions on $g_{ij}$ and $k_{ij}$ in the
definition of reflection-symmetric, LRS or isotropic for
all $t$ in some interval, but no assumption is made on $f$
then the solution will be said to have reflection-symmetric, LRS
or isotropic geometry respectively on that interval. It
follows from (2.4) that $T_{ij}$ has a corresponding symmetry
property.

In this paper only reflection-symmetric solutions of (2.1)-(2.7)
are considered. The alternative notation $a^2=g_{11}/g_{11}(t_0)$,
$b^2=g_{22}/g_{22}(t_0)$, $c^2=g_{33}/g_{33}(t_0)$ is used when it
is convenient. Yet another form of (2.6) can then be obtained by doing
a change of variables in (2.9).
$$\rho=\int f_0(aw_1,bw_2,cw_3)(m^2+\delta^{rs}w_rw_s)^{1/2}dw_1dw_2dw_3
\eqno(2.13)$$
The geometric interpretation of the $w_i$ is that they are the components
of the momentum in an orthonormal frame. Similarly:
$$T^i_i=\int f_0(aw_1,bw_2,cw_3)w_i^2(m^2+\delta^{rs}w_rw_s)^{-1/2}
dw_1dw_2dw_3\eqno(2.14)$$

In order to study the dynamics of the solutions of the system (2.1)-(2.7)
in detail it is useful to introduce certain dimensionless variables
which remain finite at the singularity. It follows from (2.1) that
$H$ never vanishes except in the case of flat spacetime, which is
excluded from consideration in the following. By replacing $t$ by $-t$
if necessary, it can be arranged that $H<0$ everywhere. It will be assumed
that this has been done. Define
$$\eqalignno{
&\hat k_{ij}=k_{ij}/H&(2.15)              \cr
&\hat\rho=\rho/H^2&(2.16)                \cr
&\hat T_{ij}=T_{ij}/H^2&(2.17)           \cr
&\tau(t)=-\int_{t_0}^t H(t') dt'&(2.18)}$$
In terms of these variables equations (2.1) and (2.4) become:
$$\eqalignno{
&-\hat k^i_j\hat k_i^j+1=16\pi\hat\rho&(2.19)       \cr
&\d_\tau \hat k^i_j=-12\pi\hat\rho(\hat k^i_j-\f13\delta^i_j)+8\pi
\hat T^i_j+4\pi(\hat k^i_j-\delta^i_j)\tr{\hat T}&(2.20)}$$
The following lemma provides some information about the range of $\tau$.

\vskip .5cm\noindent
{\bf Lemma 2.1} Let a solution of (2.1)-(2.7) with $m=0$ or $m=1$
be given which is the maximal globally hyperbolic development of initial
data on the hypersurface $t=t_0$. Then $H(t)$ is a monotonic function
defined on an interval $(t_1,\infty)$. By translating $t$ it can be
assumed that $t_1=0$. Then $\lim_{t\to 0} H=-\infty$ and
$\lim_{t\to\infty}H=0$. Moreover, $-3/t\le H(t)\le -1/t$.

\noindent
{\bf Proof} That $H(t)$ is monotonic and defined on an interval of the
form $(t_1,\infty)$ with $\lim_{t\to t_1}H=-\infty$ was shown for the
case $m=1$ in \ref{12}. Essentially the same argument applies
for $m=0$. In the latter case the coefficients of the characteristic
system are only Lipschitz instead of $C^1$ but this causes no problems.
Now it follows from (2.4) that $\f13 H^2\le\d_t H\le H^2$. Comparing the
solution with the ordinary differential equations corresponding to these
inequalities then gives the desired estimates (cf. \ref7).

\vskip .5cm\noindent
This result implies that the integral defining $\tau$ diverges as
$t\to 0$ and as $t\to\infty$. Hence the solution of (2.19)-(2.20)
exists globally in $\tau$.

The solution of (2.19)-(2.20) of course contains only a small part of
the information of that contained in the solution of (2.1)-(2.7). The
former is only a projection of the latter.  Nevertheless it will be
seen that a lot of information about the solution of the full
equations can be obtained by studying this projection. Consider now
the set $K$ of triples of real numbers $\hat k^1_1,\hat k^2_2, \hat
k^3_3$ which satisfy $\sum_i(\hat k^i_i)^2\le 1$ and $\sum_i \hat
k^i_i=1$. This is a compact subset of $\R^3$. In fact it is a disc in
a plane. A solution of (2.19)-(2.20) defines a point of $K$ at each
time $\tau$. It is on the boundary of $K$ in the plane if $f$ is
identically zero and in the interior otherwise. This point depending
on $\tau$, considered as a mapping from $\R$ to $K$, will be referred
to as the projection of the given solution. The projection of a vacuum
solution is constant.  Let $C$ denote the point $(\f13,\f13,\f13)$.
If a solution has isotropic geometry on a time interval then its
projection lies at the point $C$ during this time. Conversely, if its
projection lies at $C$ on a given time interval then it can be made to
have isotropic geometry by a time-independent rescaling of the spatial
coordinates.  Let $L_1$, $L_2$ and $L_3$ be the subsets of $K$ defined
by $\hat k^2_2=\hat k^3_3$, $\hat k^1_1=\hat k^3_3$ and $\hat
k^1_1=\hat k^2_2$ respectively.  A solution has LRS geometry on a time
interval (up to a constant rescaling of the spatial coordinates as
above) if and only if its projection lies on one of the lines $L_1$,
$L_2$ or $L_3$ during this time.  Let $L_i^+$ denote the open half of
$L_i$ which ends at the point with coordinates $(-\f13,\f23,\f23)$ or
a permutation thereof and let $L_i^-$ denote the opposite half-line,
which ends at the point with coordinates $(1,0,0)$ or a permutation
thereof. Let $V_i^+$ and $V_i^-$ denote these endpoints. Let $A_1$ be
the open region bounded by $L_2^+$, $L_3^+$ and the boundary of $K$
and let $A_2$ and $A_3$ be defined by cyclically permuting $(1,2,3)$
in this definition.  Let $B_i$ be the subset of $K$ where $\hat
k^i_i\le 0$.

The components of the metric satisfy the evolution equations
$$dg_{ii}/d\tau=2\hat k^i_i g_{ii},\ \ \ \ \ \ \ \ \ \ \ \ \
d/d\tau(g_{ii}/g_{jj})=2(\hat k^i_i-\hat k^j_j)(g_{ii}/g_{jj})\eqno(2.21)$$
which imply that $g_{ii}$ or their ratios increase or decrease
exponentially if certain sign conditions are satisfied by the
$\hat k^i_i$. There are of course corresponding statements
for the scale factors $a$, $b$ and $c$. Given an initial datum
$f_0$ for $f$ the quantities $\rho$ and $T^i_i$ are determined
uniquely by the $g_{ii}$ by means of equations (2.13) and (2.14).
Thus (2.19), (2.20) and (2.21), together with the equation
$$\d_\tau H=-H(1-12\pi\hat\rho+4\pi\tr {\hat T})\eqno(2.22)$$
derived from (2.4) form a closed system of ordinary
differential equations, which formally determine the quantities
$\hat k^i_i$, $g_{ii}$ and $H$ as functions of $\tau$ in terms
of initial data. If the coefficients of this system were locally
Lipschitz, it would follow from the standard uniqueness theorem for
ordinary differential equations that they determine them uniquely.
It will now be shown that in fact for $m=1$ this dependence is analytic.
To do this it is convenient to use the expressions for $\rho$ and
$T^i_i$ of the type (2.9). Analyticity is a consequence of the
following lemma:

\noindent
{\bf Lemma 2.2} Let $W$ be a mapping of $U\times\R^3$ to $\R$,
where $U$ is an open subset of $\R^3$. Suppose that $W$ extends to a
$C^1$ mapping $\tilde W$ of $\tilde U\times\R^3$ to $\C$, where
$\tilde U$ is an open neighbourhood of $U$ in $\C^3$, and that
$\tilde W(\cdot,y)$ satisfies the Cauchy-Riemann equations for
each fixed $y\in R^3$. Finally suppose that each $z\in \tilde W$ has
an open neighbourhood $V$ such that the supports of the functions
$\tilde W(z,\cdot)$ are contained in a common compact subset $K$
of $\R^3$. Then the function $F(x)=\int_{\R^3}W(x,y)dy$ is analytic.

\noindent
{\bf Proof} It suffices to show that the function
$\tilde F(z)=\int_{\R^3}\tilde W(z,y)dy$ is complex analytic and
this is true if $\tilde F$ is $C^1$ and satisfies the Cauchy-Riemann
equations \ref{13}. The assumptions on the smoothness and support
of $\tilde W$ justify differentiation under the integral and so
the Cauchy-Riemann equations for $\tilde F$ follow from the
Cauchy-Riemann equations for $\tilde W$.

\vskip  10pt\noindent
A consequence of the analyticity of the coefficients in the system of
ordinary differential equations is that the solutions are analytic. It
follows that if a solution with $m=1$ has LRS or isotropic geometry on
some non-empty open time interval then it must have LRS or isotropic
geometry respectively for all values of $\tau$. The same
conclusion holds if there is a sequence of times having a limit
point where the geometry is LRS or isotropic, respectively.

\vskip .5cm
\noindent
{\bf 3. The asymptotic system}

In this section a certain system of ordinary differential equations
is introduced and the qualitative behaviour of its solutions
analysed. This sytem  is used later to study the asymptotic
behaviour of solutions of the system (2.1)-(2.7). This system can
be obtained formally from (2.1)-(2.7) by replacing the $C^1$
function $f(t,v_1,v_2,v_3)$ by a measure of the form
$f(t,v_1)\delta(v_2-\bar v_2(t))\delta(v_3-\bar v_3(t)$ where
$\delta$ is a Dirac measure and taking $m=0$. Solutions
of this system can be interpreted as certain distributional
solutions of the Einstein-Vlasov system for massless particles.
These are intermediate between smooth solutions and the even
more singular solutions which are in one to one correspondence
with dust solutions. (For the correspondence between dust and
distributional solutions of the Vlasov equation see \ref{14}.) The
mathematical results which will now be derived are independent
of this interpretation.

The system of ODE's to be considered is the special case of the
equations (2.19) and (2.20) obtained by setting $\hat T^1_1=
\hat T^2_2=0$ and $\hat T^3_3=\hat\rho$. Note that, in contrast
to the general case of (2.19) and (2.20), these specialized
equations suffice to determine all unknowns occurring in them
from initial data. The explicit form of (2.20) in this case is:
$$\eqalignno{
\d_\tau\hat k^1_1&=-8\pi\hat\rho\hat k^1_1&(3.1)       \cr
\d_\tau\hat k^2_2&=-8\pi\hat\rho\hat k^2_2&(3.2)       \cr
\d_\tau\hat k^3_3&=-8\pi\hat\rho(\hat k^3_3-1)&(3.3)}$$
If initial data are chosen at some time which satisfy the condition
$\sum_i(\hat k^i_i)=1$ then the solution also satisfies it. Only
solutions with this property are considered here. It follows from
(3.1) and (3.2) that $\d_\tau(\hat k^2_2/\hat k^1_1)=0$ whenever
$\hat k^1_1\ne 0$. Let $r$ be the constant value of
$\hat k^2_2/\hat k^1_1$. Then
$$\eqalignno{
\hat k^2_2&=r\hat k^1_1&(3.4)                           \cr
\hat k^3_3&=1-(1+r)\hat k^1_1&(3.5)}$$
Substituting (2.19), (3.4) and (3.5) into (3.1) gives:
$$\d_\tau\hat k^1_1=(\hat k^1_1)^2[(1+r+r^2)\hat k^1_1-(1+r)]\eqno(3.6)$$

\noindent
{\bf Proposition 3.1} Let $(\hat k^1_1, \hat k^2_2, \hat k^3_3)$ be
a solution of (3.1)-(3.3) satisfying $\sum_i (\hat k^i_i)=1$ and
$\sum_i (\hat k^i_i)^2<1$. Define $r=\hat k^2_2/\hat k^1_1$ whenever
$\hat k^1_1\ne 0$. Then:

\noindent
(i) if $\hat k^1_1$ is zero initially it is always zero
\next
(ii) if $\hat k^1_1$ is initially (and hence always) non-zero then
$r$ is constant
\next
(iii) when $\hat k^1_1\ne0$ it is a monotonic function with
$\lim_{\tau\to -\infty}\hat k^1_1=(1+r)/(1+r+r^2)$ and
$\lim_{\tau\to\infty}\hat k^1_1=0$
\next
(iv) in that case $\lim_{\tau\to -\infty} (\hat k^3_3-\hat k^2_2)=
-r(r+2)/(1+r+r^2)$

\noindent
{\bf Proof} Statement (i) follows from (3.1). Statement (ii) has
been demonstrated above. Statement (iii) follows from (3.6). The last
conclusion is then an immediate consequence of the definitions.

\vskip 10pt\noindent
Of course analogous statements hold if $\hat k^1_1$ and $\hat k^2_2$
are interchanged since these two quantities occur symmetrically in the
hypotheses.

\vskip .5cm
\noindent
{\bf 4. Pressure estimates}

The results of this section are all variations on the theme that
if the spacetime is expanding in a certain direction then the
pressure in that direction tends to decrease. It is assumed throughout
that the geometry is reflection-symmetric. A solution of (2.1)-(2.7)
with $m=0$ satisfies $\rho=\tr T$. This condition never holds
when $m=1$ but the next lemma shows that it does hold asymptotically.

\vskip 10pt\noindent
{\bf Lemma 4.1} Suppose that some solution of (2.1)-(2.7) with $m=1$
is defined on an interval $(-\infty,\tau_1)$, with $f$ not identically
zero. Then $\lim_{\tau\to -\infty} \tr T/\rho=1$.

\noindent
{\bf Proof} It was shown in \ref7 that $\lim_{\tau\to -\infty}\rho=
\infty$. Thus the result will follow if it can be shown that
$\rho\to\infty$ implies $\tr T/\rho\to 1$. To do this choose some
radius $L>0$ and write the $\rho=\rho_1+\rho_2$ and
$\tr T=(\tr T)_1+(\tr T)_2$ where the first summand is the
integral over the region $|w|<L$ of the integrand in (2.13) or (2.14)
respectively and the second is the integral over the complementary region.
Using the fact that$(1+x^2)^{1/2}-x^2/(1+x^2)^{1/2}=(1+x^2)^{-1/2}$ it can
be seen that $\rho_2-(\tr T)_2\le (1+L^2)^{-1}\rho_2$. Hence
$$(\tr T)_2\ge{L^2\over 1+L^2}\rho_2\eqno(4.1)$$
and
$$\tr T\ge{\tr T}_2\ge{L^2\over 1+L^2}\left(\rho-(4\pi/3) L^3(1+L^2)^{1/2}
\|f_0\|_\infty\right)\eqno(4.2)$$
By choosing $L$ sufficiently large the quantity $L^2/(1+L^2)$ can
be made as close to unity as desired. For fixed $L$ the quantity
in brackets on the right hand side of (4.2) approaches $\rho$
as $\rho$ becomes large. This suffices to give the conclusion of
the lemma.

\vskip 10pt
For a given initial datum $f_0$ the equation (2.14) defines the
pressures $T^i_i$ as functions of $a$, $b$ and $c$. The following
results concern the qualitative behaviour of these functions.

\vskip 10pt\noindent
{\bf Lemma 4.2} If $f_0$ is not identically zero and $a\le C'\min \{1,b,c\}$
for some constant $C'>0$ there exists a constant $C>0$ such that
$T^1_1\ge C a^{-2}b^{-1}c^{-1}$ and $T^2_2/T^1_1\le C(a/b)^{4/3}$.
In the case $m=0$ the conclusion holds under the weaker hypothesis
that $a\le C'\min \{b,c\}$

\noindent
{\bf Proof} Let $p$ be a point of $\R^3$ where $f_0\ne 0$ whose first
coordinate $w_1$ is non-zero. Let $\delta$ be a positive number such
that $f$ is bounded below by some positive constant $\eta$ on the closed
cube $W$ of side $2\delta$ centred at $p$ and such that $w_1$ does not
vanish anywhere on this cube. Consider now the image $W'$ of the cube $W$
under the mapping $(w_1,w_2,w_3)\mapsto (a^{-1}w_1,b^{-1}w_2,c^{-1}w_3)$.
On $W$ the functions $w_2/w_1$ and $w_3/w_1$ are bounded. Under the
assumptions of the lemma they are bounded by the same constant on
$W'$. It follows that $|w_1|/(1+|w|^2)^{1/2}$ is bounded below on
any such cube by a positive constant which is independent of $a$,
$b$ and $c$ which satisfy the hypotheses of the lemma. The integral defining
$T^1_1$ can be bounded from below by the integral of the same quantity over
$W$. It follows that
$$T^1_1\ge C\eta a^{-2}b^{-1}c^{-1}\eqno(4.3)$$
and this proves the first part of the lemma. To get a lower bound for
$T^1_1/T^2_2$ the domain of integration in the definition of these two
quantities will be divided into the regions $|w_2|>R|w_1|$ and
$|w_2|<R|w_1|$, where $R$ is a positive number which will
be specified later. Corresponding to this decomposition of the domain of
integration, there are decompositions $T^1_1=T^1_{11}+T^1_{12}$ and
$T^2_2=T^2_{21}+T^2_{22}$. The volume of the region where $|w_2|>R|w_1|$
and $f(aw_1,bw_2,cw_3)\ne 0$ can be bounded by an expression of the form
$CR^{-1}c^{-1}b^{-2}$ and so $T^2_{21}\le CR^{-1}c^{-1}b^{-3}$. On the other
hand, $T^2_{22}\le R^2 T^1_{12}$. Thus
$$T^2_2\le CR^{-1}c^{-1}b^{-3}+R^2T^1_1\le (CR^{-1}(a/b)^2+R^2)T^1_1
\eqno(4.4)$$
where in the last step (4.3) has been used. Choosing $R=(a/b)^{2/3}$ gives
$T^2_2\le C(a/b)^{4/3}T^1_1$ and this proves the result for $T^2_2/T^1_1$.

\vskip 10pt\noindent
{\bf Lemma 4.3} Suppose that some solution of (2.1)-(2.7) is defined
on the interval $(-\infty,\tau_1)$, with $f$ not identically zero. If
$\hat k^1_1-\hat k^2_2\ge A$ and $\hat k^1_1-\hat k^3_3\ge A$ on this
interval for some $A>0$ then $T^i_i/T^1_1\le Ce^{4A\tau/3}$ for
$i=2,3$

\noindent
{\bf Proof} It suffices to note that under the assumptions of
this lemma there will be a time interval $(-\infty,\tau_2)$
where the hypotheses of Lemma 4.2 hold, so that (4.4) can be
applied.

\vskip 10 pt
The time derivatives of the quantities $\hat T^i_i$ cannot in general
be expressed in terms of the dimensionless quantities (2.9)-(2.11)
so as to get a closed system of ordinary differential equations.
However they can be estimated in terms of these quantities.
Note first that
$$d\hat T^i_i/d\tau=-H^{-3}dT^i_i/dt+2\hat T^i_i[1-12\pi\hat\rho
+4\pi\tr {\hat T}]\eqno(4.5)$$
Next, a change of variables in (2.7) gives in the diagonal case:
$$T^i_i=g^{ii}\int f_0(v_1,v_2,v_3)(v_i)^2(m^2+g^{rs}v_rv_s)^{-1/2}
(\det g)^{-1/2} dv_1 dv_2 dv_3\eqno(4.6)$$
Hence
$$\eqalign{
&dT^1_1/dt=(3k^1_1+k^2_2+k^3_3)T^1_1       \cr
&+g^{11}\int f_0(v_1,v_2,v_3)(v_1)^2F(v_1,v_2,v_3)
(m^2+g^{rs}v_rv_s)^{-3/2} (\det g)^{-1/2}dv_1dv_2dv_3}\eqno(4.7)$$
where
$$F(v_1,v_2,v_3)=(-g^{11}k^1_1(v_1)^2-g^{22}k^2_2(v_2)^2
-g^{33}k^3_3(v_3)^2)\eqno(4.8)$$
Note now that
$$|F(v_1,v_2,v_3)|\le (|k^1_1|+|k^2_2|+|k^3_3|)(m^2+g^{rs}v_rv_s)\eqno(4.9)$$
and so the integral in (4.7) can be bounded in modulus by $3HT^1_1$.
Putting this information into (4.5) gives the desired bound.

\vskip 10pt\noindent
{\bf Lemma 4.4} Consider a maximally extended solution of (2.1)-(2.7)
with $m=1$ and $f$ not identically zero. If $g_{ii}\to\infty$ as
$\tau\to \infty$ then $\lim_{\tau\to\infty}T^i_i/\rho=0$. If on
the other hand $g_{11}$ is bounded above and all $g_{ii}$ are
bounded below by a positive constant on an interval of the form
$[\tau_1,\infty)$ then $T^1_1/\rho$ is bounded below by a positive
constant on that interval.

\noindent
{\bf Proof} It follows from (2.13) and (2.14) that $T^1_1\le Ca^{-2}\rho$
and this proves the first statement. To get the other conclusion, choose
a cube $C_1$ as in the proof of Lemma 4.2. Then
$T^1_1\ge Ca^{-1}b^{-1}c^{-1}$ while $\rho\le Ca^{-1}b^{-1}c^{-1}$.
Hence $T^1_1/\rho\ge C>0$.

\vskip .5cm
\noindent
{\bf 5. The main results}

\noindent
{\bf Lemma 5.1} (Compactness lemma) Let a sequence of reflection-symmetric
global solutions of equations (2.1)-(2.7) be given. Then there exists a
subsequence such that $\hat k^i_j$ and $\hat\rho$ converge uniformly on
compact sets of $\R$. $\hat T^i_i$ also converges uniformly on compact
subsets (after possibly passing to a subsequence once more) and
the limiting quantities satisfy (2.19) and (2.20).

\noindent
{\bf Proof} The quantities $\hat k^i_j$ are contained in the compact set $K$
and so are, in particular, uniformly bounded. By (2.19) $\hat\rho$ is
uniformly bounded. It follows that $\hat T^i_j$ is uniformly bounded.
Equation (2.20) now shows that $\d_\tau\hat k^i_j$ is uniformly bounded.
By Ascoli's theorem there exists a subsequence such that $\hat k^i_j$
converges uniformly on the interval $[-1,1]$. Applying the theorem
again shows that this subsequence has a subsequence such that $\hat k^i_j$
converges uniformly on $[-2,2]$. Continuing in this way we obtain a
collection of subsequences indexed by a positive integer $n$ with
the properties that for the $n$th subsequence $\hat k^i_j$ converges
uniformly on $[-n,n]$ and each sequence is a subsequence of the previous one.
The diagonal sequence has the property that $\hat k^i_j$ converges
on each compact subset of the real line. By the Hamiltonian constraint
$\hat\rho$ also converges uniformly on compact sets along this
subsequence. In the diagonal case the derivatives $\d_\tau\hat T^i_i$
are bounded, as was shown in Section 4 and applying Ascoli's theorem
as before gives the remaining conclusions.

\noindent
{\bf Theorem 5.1} Let a global solution of equations (2.1)-(2.7) be given for
which $f$ is not identically zero. If at some time $\tau_1$ the projection of
the solution lies in the set $B_i$ for some $i$ then:

\noindent
(i) the projection lies in $B_i$ for all $\tau\le\tau_1$
\next
(ii) if there exists some $\tau_2>\tau_1$ such that the projection of
the solution lies in the complement of $B_i$ then it lies in the
complement of all $B_j$ for $\tau>\tau_2$.
\next
(iii) as $\tau\to -\infty$ the projection converges to a point of
the boundary which is not one of the points $V_i^-$

\noindent
{\bf Proof} Suppose without loss of generality that $i=1$. It follows from
(2.20) that
$$\d_\tau\hat k^1_1=-4\pi(\hat\rho-\tr{\hat T})(3\hat k^1_1-1)-8\pi
\tr{\hat T}\hat k^1_1+8\pi\hat T^1_1\eqno(5.1)$$
If at some time $\hat k^1_1\le0$ then the first and second terms on
the right hand side of (5.1) are non-negative while the third term is
positive. Hence $\d_\tau\hat k^1_1>0$. This implies the first
conclusion of the theorem. Moreover it means that if the projection
once leaves $B_1$ it can never reenter it. A similar statement of
course applies to any other $B_j$ and this gives (ii). To prove (iii)
note first that $\d_\tau\hat k^1_1$ is bounded below by a positive
constant as long as $\hat\rho$ is. This shows that $\liminf_{\tau\to -\infty}
\hat\rho=0$. Equation (5.1) also implies that the integral
of $\hat\rho$ on the interval $(-\infty,\tau_1]$ must be finite
so that for each $i$ the integral of the right hand side of (2.20) is
absolutely convergent. Hence each $\hat k^i_i$ tends to a limit as
$\tau\to -\infty$. By what has already been said it can only be a point
of the boundary of $K$. The monotonicity of $\hat k^1_1$ shows that this
limit cannot be one of the points $V_i^-$.

\noindent
{\bf Theorem 5.2} Let a global solution of equations (2.1)-(2.7) be given for
which $f$ is not identically zero. If at some time $\tau_1$ the projection of
the solution lies in the set $A_i$ for some $i$ then as $\tau$ decreases
either:

\noindent
(i) the projection converges to a point of the
boundary of $K$ as $\tau\to -\infty$ or
\next
(ii) it reaches $L_j^+$ for some $j$ or $C$ at a finite time before
$\tau_1$ or
\next
(iii) it stays in $A_i$ for all $\tau<\tau_1$ and it has a point of
one of the lines $L_j^+$ or the point $C$ as an accumulation point.

\noindent
{\bf Proof} Suppose without loss of generality that $i=1$. When the
projection lies in $A_1$ the inequalities $k^1_1>k^2_2$,
$k^1_1>k^3_3$ and $\hat k^1_1>1/3$ hold.  Suppose that on the time
interval $(-\infty,\tau_1)$ the inequalities $\hat k^1_1-\hat k^2_2\ge A$
and $\hat k^1_1-\hat k^3_3\ge A$ are satisfied for some $A>0$.
Then by Lemma 4.3 it follows that on this time interval $T^1_1/T^2_2$
and $T^1_1/T^3_3$ can be bounded below by a decreasing function which
tends to $\infty$ as $\tau\to -\infty$. Moreover, by Lemma 4.1
$\tr T/\rho\to 1$ as $\tau\to -\infty$. Now define a
sequence of solutions of (2.1)-(2.7) by $u_n(\tau)=u(\tau-n)$,
$\tau\in (-\infty,\tau_1)$, where $u$ denotes any of the functions which
make up the solution and $n$ is a positive integer. By Lemma 5.1 there
exists a subsequence such that $\hat k^i_j$ and $\hat\rho$ converge
uniformly on compact subsets. By the statements made above $\tr{\hat T}$
must tend to the same limit as $\hat\rho$ along this sequence. Also
$\hat T^1_1$ tends to this same limit and $\hat T^2_2$ and $\hat T^3_3$
tend to zero. Applying Lemma 5.1 again shows that the limits of these
sequences satisfy (2.19) and (2.20). Because of the values of the
limits they in fact satisfy the asymptotic system (3.1)-(3.3). The
solution of the asymptotic system obtained inherits the properties
that $\hat k^1_1\ge\hat k^2_2$ and $\hat k^1_1\ge\hat k^3_3$. The
only solutions of the aymptotic system which satisfy these
inequalities on an interval of the form $(-\infty,\tau_1)$ are
the vacuum solutions. If for some choice of subsequence this vacuum solution
is not that corresponding to the point $V_1^-$ then the projection of the
original solution must converge to that point, by Theorem 5.1. Otherwise
every subsequence of the sequence of translated solutions has a
subsequence which converges to the same solution of the asymptotic
system. Hence the whole sequence converges to this solution and the
projection of the original solution converges to $V_1^-$. In both
cases the solution of the original system converges a point
of the boundary of $K$.

It remains to consider the case where the above estimate is not
satisfied for any $A>1$. If the solution does not reach $L_j^+$
for some $j$ or $C$ in finite time then it stays in $A_i$ for all
$\tau<\tau_1$. Then it must have as an accumulation point either
a point on $L^+_j$ for some $j$, $C$ or $V_j^+$ for some $j$. In
the first two cases this gives case (iii) of the conclusion of
the theorem. In the third case the solution enters $B_j$ and so
by Theorem 5.1 case (i) of the conclusion holds.

\vskip 10pt
\noindent
In case (iii) of this theorem we can also consider a limit of translates
of the solutions whose existence is guaranteed by Lemma 5.1. If the
ratios $a/b$ and $a/c$ tended to zero as $\tau\to -\infty$ for
the original solution then by Lemma 4.2 the ratios $T^1_1/T^2_2$ and
$T^1_1/T^3_3$ would tend to infinity and the solution would belong to
case (i). Thus in case (iii) it can be assumed without loss of generality
(after possibly interchanging the indices 2 and 3) that $b/a$ is bounded
as $\tau\to -\infty$. Hence $\int_{-\infty}^{\tau_1}(\hat k^1_1-\hat k^2_2)$
is finite. Since $\d_\tau (\hat k^1_1-\hat k^2_2)$ is bounded it follows
that $\hat k^1_1-\hat k^2_2\to 0$ as $\tau\to -\infty$. Hence the solution
obtained as a limit of translates has LRS geometry. It also satisfies
$\rho=\tr T$. Information about the asymptotics of LRS solutions can
thus be used to obtain information about the asymptotics of the solutions
which fit into case (iii) of Theorem 5.2 but do not fit into case (i).

\vskip 10pt\noindent
{\bf Theorem 5.3} Let a solution of equations (2.19)-(2.20) be given
which satifies the LRS condition $k^2_2=k^3_3$. If at some time $\tau_1$
the projection satisfies $\hat k^1_1<1/3$ then either:

\noindent
(i) the projection of the solution tends to the point $(-\f13,\f23,\f23)$
as $\tau\to -\infty$,
\next
(ii) it tends to the point $(0,\f12,\f12)$ as $\tau\to -\infty$,
\next
(iii) it reaches the point $(\f13,\f13,\f13)$ at a finite time before $\tau_1$
or
\next
(iv) it tends to the point $(\f13,\f13,\f13)$ as $\tau\to -\infty$

\noindent
{\bf Proof} Suppose first that $\hat k^1_1\le 1/3-A$ for some $A>0$. Then
by Lemma 4.2 the ratio of $T^2_2=T^3_3$ to $T^1_1$ increases without limit.
Passing to a limit of translates in the familiar way gives a solution of
the equation $\d_\tau \hat k^1_1=-8\pi\hat k^1_1\hat\rho$ for which
$\hat k^1_1$ satisfies the same inequality as before. There are only two such
solutions, namely that for which $\hat\rho=0$ and that for which
$\hat k^1_1=0$. In the first of these cases the original solution
must enter the region $B_1$ and hence by Theorem 5.1 belong to
case (i) of the conclusion of the present theorem. The only way of
avoiding this is if, no matter which subsequence is chosen, the
limiting value of $\hat k^1_1$ is zero. Hence the projection of
original solution must converge to the point $(0,\f12,\f12)$. Thus the
solution belongs to case (ii) of the conclusion. Now consider the case
where there is no $A>0$ with the given property. If, despite this, the
ratio of $T^2_2$ to $T^1_1$ tends to infinity we can argue as before
to show that the solution belongs to case (i) or (ii). If, on
the other hand this ratio remains bounded then then ratio $a/b$
must remain bounded and hence if $\hat k^1_1$ remains smaller
than $1/3$ for ever then $\int_{-\infty}^{\tau_1}(1/3-\hat k^1_1)$
is finite. It then follows as in the discussion following the proof of
Theorem 5.2 that $\hat k^1_1\to 1/3$. Thus the solution either
belongs to case (iii) or case (iv).

\vskip 10pt\noindent
{\bf Theorem 5.4} Let a solution of the equations (2.1)-(2.7) with
$m=1$ and $f$ not identically zero be given. Then $\hat k^i_i\to 1/3$
and $T^i_i/\rho\to 0$ for each $i$ as $\tau\to\infty$.

\noindent
{\bf Proof} In \ref{12} it was shown that the scale factors $a$, $b$
and $c$ are bounded below by a positive constant on any interval of the
form $[\tau_1,\infty)$. Using (5.1) this statement can be strengthened.
Suppose that $\hat k^1_1$ were negative on an interval of the form
$[\tau_1,\infty)$. Then it would follow, as in the proof of Theorem
5.1, that the integral of $\hat\rho$ on this interval was finite.
But $\hat\rho$ is increasing on this interval, a contradiction. It
follows that each $\hat k^i_i$ must become zero after a finite time and
once this happens it must immediately become positive and stay positive.
Thus, for $\tau_1$ sufficiently large $a$, $b$ and $c$ are increasing.
Consider now the behaviour of the quantity $\min\{a,b,c\}$. Suppose first
that it tends to infinity as $\tau\to\infty$. Then by Lemma 4.4 the ratios
$T^i_i/\rho$ tend to zero as $\tau\to\infty$. Construct a limit of
translates as in the proof of Theorem 5.2 except that this time the
translations should be done in the opposite direction. Then the limiting
solution satisfies $\d_\tau\hat k^i_i=-12\pi\hat\rho(\hat k^i_i-\f13)$.
This is the equation which is satisfied by a Bianchi I solution
of the Einstein equations coupled to dust. It is well known and
also easy to see directly that in the case of dust each $\hat k^i_i$
converges to $1/3$ as $\tau\to\infty$. Because a convergent
subsequence can be extracted from any subsequence of the sequence
of translates by integers it follows that $\hat k^i_i\to 1/3$ for
the original solution of the Einstein-Vlasov system as well.
Next consider the case where $a$ is bounded on an interval of the form
$[\tau_1,\infty)$ while $\min\{b,c\}\to\infty$ as $\tau\to\infty$. Then
by Lemma 4.4 the ratios $T^2_2/\rho$ and $T^3_3/\rho$ converge
to zero as $\tau\to\infty$ while $T^1_1/\rho$ remains bounded
away from zero. Equation (5.1) implies that $\d_\tau\hat k^1_1$ is
bounded below by a positive constant if $\hat k^1_1<1/3$ and $T^2_2/T^1_1$
and $T^3_3/T^1_1$ are less than $1-A$ for some constant $A>0$. However this
contradicts the boundedness of $a$. Since the volume tends to infinity
as $\tau\to\infty$ at least one of $a$, $b$ or $c$ must tend to infinity.
It follows that to complete the proof we may assume without loss of
generality that $a$ and $b$ are bounded while $c$ tends to infinity.
By Lemma 4.4 $T^3_3/\rho\to 0$ while $T^1_1/\rho$ and
$T^2_2/\rho$ are bounded below by a positive constant. Now the
integral $\int_{\tau_1}^\infty\hat k^i_i(\tau)d\tau$ is finite
for $i=1,2$ and $\d_\tau\hat k^i_i$ is bounded. Hence $\hat k^1_1$
and $k^2_2$ tend to zero as $\tau\to\infty$ and $\hat k^3_3\to 1$.
But the given behaviour of the pressures shows that for
$\hat k^3_3\ge 1/3$ and sufficiently large times $\d_\tau k^3_3$
is negative, a contradiction. This completes the proof.

\vskip .5cm\noindent
{\bf 6. A compactification}

This section is devoted to a finer examination of LRS solutions
of the Einstein-Vlasov system with massless particles. For LRS
solutions with $k^2_2=k^3_3$ let $k=\hat k^1_1$, $q=b/a$,
$Q=T^1_1/\rho$. Then $\hat k^2_2=\f12 (1-k)$ and
$\hat\rho=(1/16\pi)(\f12+k-\f32 k^2)$.
Then the essential equations describing the
dynamics are:
$$\eqalignno{
\d_\tau k&=\f14 (1+3k)(1-k)(Q-k)&(6.1)       \cr
\d_\tau q&=\f12 (1-3k)q&(6.2)}$$
The quantity $Q$ can be expressed entirely in terms of $q$ and
the initial data as follows:
$$Q=q^2\left[{
\int f_0(v_i)v_1^2(q^2 v_1^2+v_2^2+v_3^2)^{-1/2} dv_1dv_2dv_3}
\over
\int f_0(v_i)(q^2 v_1^2+v_2^2+v_3^2)^{1/2} dv_1dv_2dv_3
\right]\eqno(6.3)$$
Substituting (6.3) into (6.1) makes the equations (6.1) and (6.2)
into an autonomous system of ordinary differential equations for
$k$ and $q$. Lemma 4.2 shows that
$Q(q)=O(q^{4/3})$ as $q\to 0$. This means that the system
(6.1)-(6.2) can be extended in a $C^1$ manner to the boundary
$q=0$. Moreover, $Q$ does not contribute to the
linearization of the extended system at the critical point
$q=0$, $k=0$. The eigenvectors of the linearization are directed
along the $k$ and $q$ axes with eigenvalues $-1/4$ and $1/2$
repectively. It follows (see e.g. \ref{15}) that the dynamical
system has an unstable manifold which is a curve tangent to
the $k$ axis at the point $(0,0)$. This shows that for any
initial value $f_0$ it is possible to find LRS solutions of
the Einstein-Vlasov system where the distribution function
has the initial value $f_0$ and where the quantities $\hat k^i_i$
converge to $(0,\f12,\f12)$ as $\tau\to -\infty$. Note that the
stable manifold is just the $k$ axis and so does not give rise
to any smooth solutions of the Einstein-Vlasov system. The
information about the linearization also determines the nature
of the phase portrait near the singular point and shows that
there are solutions for which $\hat k^1_1$ approaches zero
but turns back before reaching it. A typical feature of Bianchi
models is that the matter becomes dynamically negligeable near
the singularity. No attempt will be made here to make this
notion precise but one aspect of it is that the projection of
the solution should tend to a point of the boundary of $K$ as
$\tau\to -\infty$. The solution whose existence has just been
shown is an exception to the rule. For anisotropic Bianchi I models
with a perfect fluid no exceptional solutions of this kind exist.
However they do occur for other Bianchi types\ref{16}.

The critical points of the system (6.1)-(6.2) in the region
where $0<q$ and $-1/3<k<1$ are the points of the form
$(1/3,q_0)$ where $q_0$ has the property that $Q(q_0)=1/3$.
Differentiating (6.3) and estimating the result in an
elementary way leads to the inequality $qQ'\ge Q(1-Q)$.
This shows that the function $Q$ is strictly increasing for
$q>0$. Taking account of the limiting values of $Q$, it
follows that there is precisely one value $q_0$ for which
$Q(q_0)=1/3$. Moreover, at this point $qQ'\ge 2/9$. The
eigenvalues of the linearization at the corresponding
critical point are $-\f16\pm\f12\sqrt{\f19-2q_0Q'(q_0)}$.
They both have negative real parts and the critical point
is a sink. In particular no solution emerges from this
critical point. Thus it is seen that if the projection of
any LRS solution for massless particles approaches the point
$C$ as $\tau\to -\infty$ then the projection must stay at
$C$ for all time i.e. the solution must have isotropic
geometry. This should be compared with the results of
Newman\ref{17} on isotropic singularities in solutions of
the Einstein equations coupled to a radiation fluid.

\vskip .5cm\noindent
{\bf Appendix: Fluids with nonlinear equation of state}

Consider a perfect fluid with equation of state $p=f(\rho)$
which satisfies the following general assumptions:

\noindent
(i) $f$ is a continuous function from $[0,\infty)$ to itself
with $f(0)=0$ which is $C^1$ for $\rho>0$
\next
(ii) $0\le f'(\rho)\le 1$ for all $\rho>0$
\next
(iii) there exists a constant $C<1$ such that $p\le C\rho$
for $\rho<1$

\vskip 10pt\noindent
Assumptions (i) and (ii) are standard. The third assumption
is, when (i) and (ii) are satisfied, equivalent to the
assumption made in [7] that the solution is not asymptotically
stiff at low densities. In the case of a linear equation of
state $f(\rho)=k\rho$ the assumptions (i)-(iii) are satisfied
if and only if $0\le k<1$. In a Bianchi I spacetime it follows
from the momentum constraint (2.2) that the four-velocity of
the fluid is orthogonal to the hypersurfaces of homogeneity.
Hence the energy density $\rho$ measured by an observer
whose wordline is orthogonal to the hypersurfaces of
homogeneity is the same as that measured by a comoving
observer. Equations (2.19) and (2.20) are valid as in
the case of the Einstein-Vlasov system. For a fluid it can
be assumed without loss of generality that the solution is
reflection-symmetric because given any initial data, it
suffices to do a linear transformation of the coordinates
which simultaneously diagonalizes the metric and second
fundamental form in order to transform the given data to data
for a reflection-symmetric spacetime.

In the case of a fluid $T^i_j=p\delta^i_j$ and hence
$\hat T^i_j=\hat p\delta^i_j$, where $\hat p=p/H^2$. If
the equation of state is linear then $\hat p$ can be
expressed as a function of $\hat\rho$ alone and (2.19)
and (2.20) then reduce to a system of ordinary differential
equations which suffice to determine $\hat k^i_j$ from
initial data. For a nonlinear equation of state this is no
longer the case. The equations (2.19) and (2.20) no longer
form a closed system and must instead be considered as
the projection of a bigger system, as in the case of the
Vlasov equation. This is one reason why the linear case
has been studied preferentially in the literature.
Nevertheless it turns out that the projection can be
analysed very effectively in the general case.

The first question which needs to be addressed is that of
global existence in $\tau$ i.e. the equivalent of Lemma 2.1
for a fluid. This follows from the results of [7]. The
assumption (iii) has been used at this stage. A direct
calculation shows that the quantity $(\hat k^1_1-\hat k^2_2)/
(\hat k^1_1-\hat k^3_3)$ is independent of $\tau$ whenever
$\hat k^1_1-\hat k^3_3$ is non-zero. Moreover if $\hat k^1_1
-\hat k^3_3$ is zero at some time it remains zero. Hence
the projection of each solution is constrained to move on a
straight line in $K$ passing through the centre $C$. This
is already a much stronger statement than could be proved
in the case of the Vlasov equation. To find out how the
projection moves on this straight line, the time derivative
of the dimensionless version of the density will be calculated.
For a fluid it is given by
$$\d_\tau\hat\rho=(\hat\rho-\hat p)(1-24\pi\hat\rho)\eqno({\rm A}
1)$$
Noting that the Hamiltonian constraint implies that
$24\pi\rho\le 1$ with equality only at the point $C$, it can now
be seen that the projection moves monotonically from the boundary
of $K$ at $\tau=-\infty$ to the centre $C$ of $K$ at $\tau=\infty$.
This qualitative behaviour is independent of the equation of
state satisfying (i)-(iii). The only difference is in the speed
with which the projection moves along the radial line at different
times. Equation (A1) also makes clear that this picture changes
completely if the equations of state considered here are replaced
by the limiting case of a stiff fluid, $p=\rho$.

\vskip .5cm
\noindent
{\bf References}

\noindent
1. Wald, R. M.: General Relativity. (Chicago University Press, 1984)
\next
2. Jacobs, K. C. {\it Astrophys. J.} 153, 661 (1968).
\next
3. Collins, C. B.: More qualitative cosmology. {\it Commun. Math.
Phys.} 23, 137-158 (1971).
\next
4. Coley, A. A., Wainwright, J.: Qualitative analysis of two-fluid
Bianchi cosmologies. {\it Class. Quantum Grav.} 9, 651-665 (1992).
\next
5. Collins, C. B.: Qualitative magnetic cosmology. {\it Commun.
Math. Phys.} 27, 37-43 (1972).
\next
6. Leblanc, V. G., Kerr, D. and Wainwright, J.: Asymptotic states
of magnetic Bianchi VI${}_0$ cosmologies. {\it Class. Quantum Grav.}
12, 513-541 (1995).
\next
7. Rendall, A. D.: Global properties of locally spatially homogeneous
cosmological models with matter. Preprint gr-qc/9409009 (to appear in
{\it Math. Proc. Camb. Phil. Soc.})
\next
8. Lukash, V. N., Starobinski, A. A.: The isotropization of the
cosmological expansion owing to particle production. {\it Sov.
Phys. JETP} 39, 742-747 (1974).
\next
9. Ehlers, J., Geren, P. and Sachs, R. K.:  {\it J. Math. Phys.} 9,
1344 (1968).
\next
10. Rein, G.: Cosmological solutions of the Vlasov-Einstein system
with spherical, plane and hyperbolic symmetry. Preprint gr-qc/9409041
(to appear in {\it Math. Proc. Camb. Phil. Soc.})
\next
11. Maartens, R., Maharaj, S. D.: {\it Gen. Rel. Grav.} 22, 595 (1990).
\next
12. Rendall, A. D.: Cosmic censorship for some spatially homogeneous
cosmological models. {\it Ann. Phys.} 233, 82-96 (1994).
\next
13. Griffiths, P., Harris, J.: Principles of Algebraic Geometry.
Wiley, New York, 1978.
\next
14. Rendall, A. D.: Cosmic censorship and the Vlasov equation.
{\it Class. Quantum Grav.} 9:L99-L104 (1992).
\next
15. Hartman, P.: Ordinary Differential Equations. Birkh\"auser,
Boston, 1982.
\next
16. Wainwright, J., Hsu, L.: A dynamical systems approach to
Bianchi cosmologies: orthogonal models of class A. Class. Quantum
Grav. 6, 1409-1431 (1989).
\next
17. Newman, R. P. A. C.: On the structure of conformal singularities in
classical general relativity. {\it Proc. R. Soc. London}
A443, 473-492; 493-515 (1993).

\end